\documentstyle[preprint,aps]{revtex}
\begin{document}
\draft
\title {Low energy and dynamical properties of a single hole in the
$t-J_z$ model}
\author{Q.F. Zhong and S, Sorella}
\address{International School for Advanced Study, Via Beirut 4, 34013
Trieste, Italy}
\date{\today}
\maketitle
\begin{abstract}
We review in details a recently proposed technique to extract
information about dynamical correlation functions of    many-body
hamiltonians with a few Lanczos iterations
  and without the limitation of finite  size.
 We apply this technique
to understand the low energy properties and the dynamical spectral weight of a
simple model describing the motion  of a single hole in a quantum
antiferromagnet: the $t-J_z$ model in two spatial dimension and for a double
chain lattice. The simplicity of the model allows us a well controlled
numerical
solution, especially for the two chain case.
Contrary to previous approximations
we have found that the single hole ground state in the infinite system
is continuously connected with the  Nagaoka fully polarized state
 for $J_z \to 0$.
 Analogously we have obtained an accurate determination of
the dynamical spectral weight relevant for photoemission experiments.
For $J_z=0$ an  argument is given that the
spectral weight vanishes at the Nagaoka energy
  faster than any power law, as supported also
 by a clear numerical evidence.
It is also shown that
spin charge decoupling is an exact property for a single hole
in the Bethe lattice but does not apply to
the  more realistic lattices where the hole can describe
closed loop paths.
\end{abstract}
\narrowtext
\section{Introduction}
After the discovery of the high-$T_c$ materials\cite{bm},
there have been an increasing attention
to the study of strongly correlated fermion systems. In his pioneer
work \cite{anderson} Anderson first pointed out
that a doped Mott insulator, described by the one band Hubbard model,
should contain the basic physics of the high-$T_c$ superconductors.
The Hubbard model was proposed independently in 1963
by Gutzwiller, Hubbard and Kanamori\cite{gutzwiller,hubbard,kanamori}.
Despite of its simple-looking hamiltonian
and a lot of efforts, the physics of the model is  still
a subject of debate.

In the large $U$ limit, it is more convenient to use a canonical transformation
to project out the doubly occupied sites, costing energy $U$.
This leads to the so-called $t$-$J$ model:
\begin{equation}\label{tj}
H=-t\sum_{<i,j>,\sigma} (c^\dagger _{i\sigma}c_{j\sigma} + h.c. ) +
J \sum_{<i,j>} ({\bf S}_i \cdot {\bf S}_j - {1 \over 4} n_i n_j ).
\end{equation}
where the constraint of no double occupancy is understood.
Here $c_i^\dagger$($c_i$) creates (annihilates) an electron at site $i$,
$n_i=\sum \limits_\sigma n_{i \sigma}$ is the corresponding  density
operator  with $n_{i\sigma}=c^\dagger _{i\sigma} c_{i\sigma}$,
the symbol $<i,j>$ means summation over nearest neighbors, $J={4 t^2 \over U}$
 is the superexchange coupling  and finally
the spin density operator ${\bf S}_i$ is defined by the Pauli matrices ${\bf
\vec \sigma} $: ${\bf S}_i =
\sum\limits_{\sigma,\sigma^\prime} c^\dagger_{i,\sigma} {{\bf \vec
\sigma}_{\sigma,\sigma^\prime} \over 2} c_{i,\sigma^\prime}$.

Starting from a more realistic three band Hubbard hamiltonian describing
the Copper-Oxygen layer, Zhang and Rice showed that
the low-energy physics of cuprate superconductors is determined by the
singlet state formed by the additional hole on oxygen with the existing hole
copper.\cite{zhang}
They also  showed that
the hopping of this singlet is described by an effective one-band
$t$-$J$ model,  with $J$ and $U$ essentially unrelated.
 This makes  the $t-J$ model even more appropriate to describe the Copper-Oxide
planes at low energies rather  than the original one-band Hubbard model.


During the last few years, a huge amount of analytical
and numerical work has been devoted to the study of the $t-J$  model.\cite{yu}
However the physics contained in
the $t$-$J$ model is far from clear because of the interplay between
antiferromagnetic long range order and charge degrees of freedom.
Exact numerical methods are usually limited to a very small linear size  in
more
than 1D and practically  nothing about the low energy physics
has been understood numerically.

In this paper we will make  a detailed numerical study of a simplified
version of the $t$-$J$ model, {\it i.e.} neglecting the spin fluctuations
in the exchange term (${\bf S}_i \cdot {\bf S}_j \to S^z_i S^z_j$ in Eq.
(\ref{tj})) and considering only the properties of a single hole.
Due to the simplification of the model, we are able to work directly in the
infinite size system and have a satisfactory description
 for the low energy dynamics of a single hole in such a simple model of an
antiferromagnet:
 \begin{equation}
H=-t\sum_{<i,j>,\sigma} (c^\dagger _{i\sigma}c_{j\sigma} + h.c. ) +
J_z \sum_{<i,j>} S_i^z S_j^z.
\label{tjz}
\end{equation}
 A basic motivation of this  work is to obtain
  reliable numerical results on this simple $t-J_z$ model since it
may be useful for further investigation of the more interesting $t-J$ model.
For instance, the $t-J_z$ model was often used in the past\cite{trugman,klr}
to test several approximations on the $t-J$ model.
In fact most of the physical properties of the $t$-$J$ model probably remain
valid even for  the $t$-$J_z$ model.
{}From this point of view it is worth
mentioning that the $t-J$ model and the $t-J_z$ model  have the same limit for
$J$ or $J_z \to 0$ and for infinite spatial dimensions
 the two model coincide, since the spin fluctuations are irrelevant in this
 limit.\cite{vollhardt}

The single hole problem represents certainly a further simplification but
is still physically relevant, since the single particle excitations in
magnetic insulators can actually be studied by the photoemission and the
inverse photoemission spectroscopy.

For a single hole a rigorous theorem proven by Nagaoka is known for $J_z=0$.
The so called Nagaoka theorem states that
 for a bipartite finite lattice  in more than one dimension ($d >1$)
the ferromagnetic state with maximum spin is the unique ground state
with energy $e_N=-z t$,
in any subspace with given total spin projection.

 A complete description of
the one hole spectrum in the $J_z=0$ limit,  not limited  to the ground
state, was first given in \cite{br}, where the
so-called ``retraceable path approximation'' (RPA) was introduced. In
 the Ising limit  as a hole hops in a
N\'eel state, it  scrambles the spins along its path. In order to return
the spin configuration to its original state,
it was argued that, to a good approximation, one can consider only paths
in which  the hole retraces its path back to the origin, thereby returning all
of the  spins to their original position. In this
approximation one can write down an explicit analytic solution
for the Green's function:
\begin{equation}
G_k(\omega)={z \sqrt{\omega^2-4(z-1)}-(z-2)\omega \over
2(\omega^2-z^2)}
\label{greenbr}
\end{equation}
and the spectral weight
$A(k,\omega)=-{ {\rm sgn}\,\, \omega  \over \pi} Im G_k(\omega)$ reads
\begin{equation}
A(\omega) = { z \sqrt{ 4 (z-1) t^2 - \omega^2} \over 4 \pi (z^2 t^2 -
\omega^2)}
\label{spectralbr}
\end{equation}
where $z=2d$ is the  number of nearest neighbors.
The spectral weight is completely incoherent and
dispersionless with a one particle band,
which is $75\%$ narrower (in 3D) than the noninteracting band.
The RPA is exact in 1D and for the Bethe lattice (where
Nagaoka theorem does not apply),
and recently it has been shown to be exact
in the limit of infinite spatial dimensionality. \cite{vollhardt}

For finite $J_z/t$,  the competition between
the kinetic energy $t$ favoring the ferromagnetic configuration and
the exchange energy $J_z$ favoring the antiferromagnetic alignment of the
neighboring spins makes the problem of particular interest.
By neglecting closed loops, e.g. in the Bethe lattice case,
it is possible to derive a closed solution\cite{russia,shraiman},
which we will refer in the following as the  ``string picture''.
In the string picture,
the hole moves in an antiferromagnetic spin background, leaving behind a
string of overturned spins, which costs an energy proportional to the
length of the path. The overturned spins behave like an effective linear
potential for the hole. In the continuum limit, valid
for $J_z \to 0$, the problem is reduced to a
one dimensional Schr\"odinger equation
with a linear potential $V \sim x J_z$ where $x$ is  the length of the string.
\begin{equation}
H = - \sqrt{z-1} t {\partial^2 \over \partial x^2} + {J_z(z-2) \over 2}
 x - 2 \sqrt{z-1} t.
\label{string}
\end{equation}
The solution of this hamiltonian leads to a series of bound states. The hole
is essentially localized with a discrete energy spectrum:
\begin{equation}\label{stren}
E_n = a_n \sqrt{z-1} t ( { J_z  (z-2) \over 2 \sqrt{z-1} t})^{2/3} - 2
\sqrt{z-1}
t  \end{equation}
where $a_n$ are the zeros of the Airy function $Ai(z)$.
Recently, it has been shown that
 the string
picture is exact up to order ${1 \over d^2}$,
where $d$ is the spatial dimensionality.\cite{vollhardt}

Contrary to the $J_z=0$ case, at finite $J_z$ the spectral weight has
$\delta-$function peaks at energies $E_n$. The weight of the $\delta-$
function at the lowest energy is called the quasiparticle weight $Z$
and is found to vanish linearly in $J_z$, using the continuous limit
(\ref{string}).\cite{klr}

Later Kane, Lee and Read\cite{klr} introduced a self-consistent Born
approximation that can be considered  an extension of the  RPA
to the  more  physical $t-J$  model.
 Their approach is widely accepted since they were able to
reproduce the asymptotic behavior of the $t-J_z$ model in the string picture
and a large amount of numerical work
on small lattices seems to be in qualitative agreement
with the predictions of this theory.\cite{poilblanc,dagotto,prelovcek}

In the retraceable path approximation or  in the string picture and similarly
within the self-consistent Born approximation,
 closed loop paths are neglected in order to simplify the analytic structure of
the one hole Green's function.  However, this certainly introduce
an approximation
even in the simple $t-J_z$ model.
In fact the most important effect due to the inclusion of closed loop paths in
this  model was first noted by  Trugman.\cite{trugman}  In
2D a hole can hop  around a square plaquette one and half times without
disturbing the  spin background with a net translation  to the
next-nearest-neighbor along the diagonal.  This means that the hole can
``unwind'' the string and self-generate a  next-nearest-neighbor hopping.
The full localization of charge
carriers found within the retraceable path approximation
and the string picture is an artifact of the approximation.

Another interesting issue is whether the spin and charge degrees of freedom
are decoupled or not.\cite{putikka,ogata}
Spin charge decoupling is a well known phenomenon
in one dimension, where
a one electron excitation can be decomposed into a spinon excitation which
carries spin but no charge and a holon excitation which carries charge
but no spin.
We will  show in the present paper that spin charge decoupling
is an {\em exact} property of the $t$-$J_z$ model in the Bethe lattice
with arbitrary coordination number  $z$ at single hole doping
and present some numerical work, ruling out the possibility of spin charge
decoupling in physically relevant lattices where closed loops are allowed.

In this paper, we make use of  the Lanczos  method  to calculate the single
hole spectral function in the infinite system. We will first
introduce a transformation to eliminate the charge degree of freedom
to get an effective spin Hamiltonian in
a given momentum subspace (Section 2).
Then we will use the Lanczos method in order
to diagonalize the effective spin Hamiltonian
in the infinite lattice (Section 3).
We will concentrate on two chains (2C) and two-dimensional (2D) square
lattices. The 2C case is much easier  for numerical study compared to the  2D
lattice.
The 2C lattice is not a trivial lattice for
the $t$-$J_z$ model, because it has the basic properties of
 an higher dimensional lattice, such as the existence
of closed loop paths and the validity of the Nagaoka theorem.

Most of the above results were obtained  by use of  the Lanczos Spectra
 Decoding method. This method   allows us
to analyze the Lanczos data in the infinite system,
where only a small number of
Lanczos iterations is possible (Section 3).
The validity of the Lanczos Spectra Decoding is proved analytically and
numerically  on the Bethe lattice (Section 4).
In section 5, we will show  our numerical results both for the $J_z=0$ and
finite $J_z$ case.
Finally  we will discuss about the formation of a ferromagnetic polaron around
the hole for small $J_z$, and about the analytic form of the spectral weight
close to the Nagaoka energy.

\section{Formalism}
\subsection{Effective Spin Hamiltonian}

 The $t-J_z$ Hamiltonian (\ref{tjz})
is translational  invariant, so  the most general one-hole
state  with total spin $S^T_z=M_z+{1 \over 2}$
and total momentum $-p$ ( hole momentum $p$ ) can be written:
\begin{equation}\label{state}
|\psi_p> \,=\,{1\over \sqrt N } \sum \limits_{R,\sigma}
e^{i p R} c_{R,\sigma} T_R  |S_O>
\end{equation}
where $|S_O>$ is a pure spin state that satisfies:
i) $n_R |S_O> = |S_O>$ for all sites $R$,  ii)
$n_{\downarrow O} |S_O> \,=\, |S_O>$ {\it i.e.} the spin at the origin is
fixed  to $\sigma_0 =- {1 \over 2}$, iii) the total spin along the $z$-axis
is a well defined quantum number in the state $|S_O>$,
{\it i.e.} $S_z^T |S_O> = M_z |S_O>$.
The operator $T_R$ in (\ref{state})
is the translation operator that brings the origin $O$
to the lattice point $R$. It is formally defined by:
\begin{equation}\label{translation}
T_R c_{R^{\prime},\sigma} T_{-R} \,=\, c_{R+R^{\prime},\sigma}
\end{equation}
States with definite momentum  $|\psi_p>$ are eigenstates of $T_R$
such that $T_R |\psi_p>\,=\, e^{-i p R} |\psi_p>$.
Note also that the sum over $\sigma$ in the definition of $|\psi_p>$
in (\ref{state})
is used only for later convenience.



A pure spin Hamiltonian can be derived as following.
We evaluate the expectation value of the $t$-$J_z$ Hamiltonian
on the one-hole state $|\psi_p>$, $E_p \,=\, < \psi_p| H |\psi_p>$, and
consider first the case $J_z=0$:
\begin{equation}
E_p \,=\, {t\over N } \sum \limits_{R_1,R_2,R,\tau_{\mu},
\sigma, \sigma_1, \sigma_2} e^{-i p (R_1-R_2)}
 <S_O| T_{-R_1} c^{\dag}_{R_1, \sigma_1}
\hat P \,\,\, c_{R+\tau_{\mu}, \sigma} c^{\dag}_{R,\sigma}
 c_{R_2,\sigma_2} T_{R_2}|S_O>
\end{equation}
Now i) $\sigma_1\,=\,\sigma_2$ since the total $S_z^T$ has to be conserved
in order to give a non-vanishing contribution in $E_p$.
ii) $R=R_2$ and $R+\tau_{\mu}=R_1$, otherwise we create a doubly
occupied site which is projected out  either by $ \hat P$ or by
overlapping with $<S_O|$.
Finally we obtain that $E_p\,=\, <S_O| H^{eff}_p |S_O>$ with:
\begin{equation}\label{effha}
H^{eff}_p=t \sum \limits_ {\sigma,\sigma^{\prime}, \tau_{\mu}}
e^{-i p \tau_{\mu}} T_{-\tau_{\mu}} (c^{\dag}_{\tau_{\mu},\sigma^{\prime}}
c_{\tau_{\mu},\sigma} c^{\dag}_{O,\sigma} c_{ O, \sigma^{\prime}}) .
\end{equation}
$H^{eff}_p$ acts only on spin states. In fact for $S={1\over 2}$
it is possible to express the term between parenthesis in terms of
spin operators only:
\begin{equation}
\chi_{R_i,R_j}= \sum\limits_{\sigma,\sigma^{\prime}}
 c^{\dag}_{R_i,\sigma^{\prime}} c_{R_i,\sigma}
c^{\dag}_{R_j,\sigma} c_{R_j, \sigma^{\prime}}\,=\,
{1\over 2} n_{R_i} n_{R_j} + 2 {\bf S}_{R_i} \cdot {\bf S}_{R_j}
\end{equation}
Finally $H^{eff}_p$ reads:
\begin{equation}
\label{heff}
H_p^{eff} = t \sum \limits_{\tau_{\mu}} e^{-i p \tau_{\mu}}
 T_{-\tau_{\mu}} \chi_{O,\tau_{\mu}}
=t \sum \limits_{\tau_{\mu}} e^{i p \tau_{\mu}}
  \chi_{O,\tau_{\mu}} T_{\tau_{\mu}}.
\end{equation}
Note that the product of the operators  $ T_{-\tau_{\mu}} $
and  $ \chi_{O,\tau_{\mu}}$ appearing in $H^{eff}_p$ permutes
spins and leaves  the origin $O$ unchanged.
Moreover, since  the operator $\chi$  interchanges the spins at sites $R_i$
and $R_j$, we have that $\chi_{R_i,R_j}= \chi_{R_j,R_i}$  and that
$\chi_{R_i,R_j}^2=1$.

Analogoously we can extend to  $J_z \ne 0$ the derivation  of
 a pure spin Hamiltonian and obtain:
\begin{equation}\label{eff}
H^{eff}_p \,=\, t  \sum \limits_{\tau_{\mu}}  e^{i p \tau_{\mu}}
 \chi_{O,\tau_{\mu}}T_{\tau_{\mu}}   \,+\, J_z \sum \limits_{(i,j)\ne O}
S^z_{R_i} S^z_{R_j}
\end{equation}
The full Hamiltonian commutes with ${\bf S}_O$ (since it actually
does not depend on the spin at the origin) and thus it can be defined on
$N-1$ sites.


In conclusion for any eigenstate  $|S_O>$ of $H_p^{eff}$ with definite $S^z_O
|S_O> = -{1 \over 2} |S_O>$ we have an eigenstate of the $t-J_z$ Hamiltonian
written in the form (\ref{state}).
In fact, by use of the variational principle any eigenstate of (\ref{tjz})
or (\ref{eff}) is obtained by ${\delta \over \delta \psi_p} <\psi_p | H |
\psi_p > = 0 $, ${\delta \over \delta S_O} <S_O | H |
S_O > = 0 $
 with the condition $<\psi_p | \psi_p>=1$, $<S_O|S_O>=1$ respectively.
Since $E_p
= < \psi_p | H | \psi_p > = < S_O | H_p^{eff}|S_O>$ and
$<\psi_p|\psi_p>=<S_O|S_O>$, it  clearly follows
that all the eigenstates of $H_p^{eff}$ with definite spin at the origin define
a true eigenstate of $H$ by use of  (\ref{state}).
In conclusion the one hole problem is mapped  onto
a diagonalization of a pure spin Hamiltonian
$H_p^{eff}$ for the given momentum $p$. \cite{xiang}
A similar effective Hamiltonian can be obtained for the $t$-$J$ model,
by substituting $S^z$ with ${\bf S}$ in (\ref{eff}).

It is interesting to note that $H_p^{eff}$ is not translational
invariant and that the momentum of the hole appears as a simple parameter.
It is just  this property that allow us to diagonalize the
$t$-$J_z$ model in certain momentum subspace in the infinite lattice.

As it is shown in the Appendix many useful dynamical quantities
such as the Green's function and current operators can be easily translated
in terms of spin operators acting on this spin space in which the hole
is fixed at the origin.

\subsection{Green's Function for $J_z = 0$}

Following Brinkman and Rice\cite{br}, we can expand the Green's
function $G$ in terms of the momenta  $<H|(H_p^{eff})^k|H>$  of the Hamiltonian
on the  translation invariant ground state of the undoped system $|H>$.
For $J_z=0$, at vanishing doping, the hamiltonian is classical and $|H>$ is
given by:
  \begin{equation}
|H>={1\over \sqrt{2}} (|N> + |N^{\prime}>)
\label{defh}
\end{equation}
 where $|N>$ and
$|N^{\prime}>$ are  the two possible determination  of the N\'eel state.

The one hole Green's function can be generally
written as a summation of all possible paths traced by the hole
during its motion on the lattice. A path is then defined by a set of
coordinates $\{R_l\}^n$ with $l=0,...,n$,  which are connected by nearest
neighbor vectors $\tau_{\mu}$ \begin{equation} R_l-R_{l-1} = \tau_{\mu_l}.
\end{equation} Among  all  the possible  paths
it is useful to identify   the skeleton ones. \cite{mh}
A  path $\{R_l\}^n$ of length $n$ is a skeleton path if
$R_{l+1} \not= R_{l-1}$ for any $l=1,...,n-1$. By definition, for any skeleton
path,  the hole never retraces its path immediately. It is clear that
all  the remaining  paths
can be obtained by dressing each site $R_l$ of the skeleton path by all
possible
retraceable paths.
The retraceable paths can be then summed exactly.

Instead of giving the detailed derivation, we will present the final result
and discuss some of the special cases.
The most general Green's function can be written in a formal expansion
of skeleton paths of length $2 n$:
\begin{equation}\label{green}
 G(R, \omega)\,=\, G_{BR} ( \omega)
\sum \limits_n K( \omega)^{2n} C_n(R)
\end{equation}
where $G_{BR}$ is the Brinkman Rice result (\ref{greenbr}), which includes only
retraceable path contribution starting  from the origin $R=O$ and coming back
to the same site:
\begin{equation}
\label{BR}
  G_{BR}( \omega) \,=\, {1\over  \omega \left(
1- {z\over z-1}
 \Sigma^A ( \omega) \right)  }  \,=\,t^{-1}
{ K( \omega) \over 1- K( \omega)^2 },
\end{equation}
while the function $K(\omega)$
is the exact contribution of all possible  retraceable paths on
 each site of the skeleton path different from the origin:
\begin{equation}\label{nonrelevant}
K( \omega) =  {t \over   \omega} {1 \over 1- \Sigma^{A}( \omega)}
\,=\, {1\over 2(z-1) t} \left[  \omega -{\rm sgn} (\omega)
\root \of {  \omega^2 -
4 (z-1) t^2 } \right ]
\end{equation}
Finally in (\ref{green}) the coefficients $C_n(R)$ are determined by:
$C_n(R)=\sum\limits_{all\,\,\, skeleton
\,\,\, \{R_l\}^{2 n} } \Omega(\{R_l\}^{2 n}) $, where
 $\Omega(\{R_l\}^n)$ are spin correlation
functions, which, generally speaking, depend on the path,
\begin{equation}
\Omega (\{R_l\}^n) \,=\, <H| \chi_{R_n,R_{n-1}} \chi_{R_{n-1},R_{n-2}}
\dots \chi_{R_1,R_0} |H>
\end{equation}
In the Ising case $\Omega (\{R_l\}^{n}) $ vanish for odd $n$ and
are simply equal to either zero or one for even $n$
depending on whether the skeleton path change the N\'eel order or not.
Instead they are complicated correlation
functions in the $t-J$ model for $J \to 0$.  In this case the expression
(\ref{green}) is still valid provided we include the odd $n$ contributions.

Two possible limits can be  exactly solved using the previous expression
(\ref{green}) for the Green's function.

\subsubsection{Bethe Lattice Case }
The Bethe lattice is defined on a Cailey-tree with coordination $z$
and for $z=2$ it coincides with the one dimensional chain.
In the Bethe lattice
there is only
one path that connects  two arbitrary sites of the lattice because
 there are no closed  loops in this lattice (see e.g. \cite{prelovcek}
for a more detailed definition of the Bethe lattice).
 As in one dimension there is only one  skeleton path connecting the origin
to a given point $R$ and  $\Omega(\{R_l\}^n )\,=\, \Omega (R_n) $,  which is
only a function of the final position $R_n$.  We can immediately write the
exact expression of the Green's  function on a Bethe lattice using the previous
general expression (\ref{green}):
\begin{equation}\label{bethegreen}
G(R_n,\bar \omega) = \Omega(R_n) \,\, G^{Free}(\bar \omega)
\end{equation}
where $G^{Free}=G_{BR}(\bar \omega) \,\, K(\bar \omega)^n$.
It is interesting that the ``strong correlation'' in the one hole problem
in the Bethe lattice is only contained in the static function
$\Omega (R_n)$,
since $G^{Free}(\omega)$ is exactly the free electron Green's function.
This has direct consequences on the spin charge
decoupling. (See next section)

\subsubsection{ $t$-$J_z$: Brinkman-Rice Almost ``Exact''}

Let us consider the diagonal Green's function $G(R=O,\omega)$ for the
$t$-$J_z$ model.
In the general expression (\ref{green}),
$\Omega(\{R_l\}^n)$ is either one or zero depending on whether a given
permutation  preserves   the N\'eel order or not.
The shortest skeleton paths with $\Omega(\{R_l\}^n) =1$ is known to be
three times the path  around the elementary square plaquette.
Thus $n=12$ for such a skeleton path.
There are $8$ possibilities to build such a path starting from the origin
in 2D ( 4 neighbors times 2 possible opposite directions) and
$  2 d ( 2 d-2)$  in dimension $d$.
Then the next leading correction to the Brinkman and Rice
result is
\begin{equation}
G(R=O,\omega) \,=\, G_{BR}(\omega) \left( 1+ 2 d (2 d-2) \,\, \displaystyle
 K(\omega)^{\displaystyle 12} + ... \right)
\end{equation}
Since $|K(\omega)|\le {1\over \root \of {2 d-1}}$ for any $\omega$,
the correction
to the Brinkman and Rice result turns out to be less then $1.1 \%$ in $d=2$
and $0.15 \%$ in $d=3$.
This has also been noted in ${1 \over d}$ expansion\cite{vollhardt}
but appears also quite natural in this formalism.

\subsection{Spin Charge Decoupling}

In 1D, if spin charge decoupling occurs, the one particle
Green's function can be written as a simple product of a spinon
contribution and a holon contribution:
\begin{equation}\label{scdc}
G(R,t)= G^{spinon}(R,t)\,\, G^{holon}(R,t)
\label{greens}
\end{equation}
This property is exact in $d=1$ for the infinite $U$ Hubbard model and is
asymptotically valid for large $(R,t)$ at finite $U$.\cite{parola}

The spin charge decoupling manifests itself in the one particle  Green's
function and in principle can be  detected  even at higher
dimensionality as speculated by several authors  following  P.W.
Anderson.\cite{anderson}
For $J, J_z \to 0$, or $U \to \infty$  one
expects no dynamics for the spinons and that the holon contribution
has exactly the free particle behavior, because there is only a unitary charge
carried by the single hole.

Thus, as a   consequence of spin charge separation,
the  one hole Green's function should  be written in the following way:
\begin{equation}
G(R,t) = \Omega(R) \,\, G^{Free} (R,t)
\label{greensep}
\end{equation}
The free electron Green's function is nothing but the free propagator in the
Nagaoka limit:
\begin{equation}
G^{Free}(k,\omega)={ 1 \over \omega - \epsilon_k + i \delta {\rm sgn} \,
\omega}
\end{equation}
where $\epsilon_k=2t ( \cos k_x + \cos k_y ) $ is the energy of a free hole.

By Fourier transforming in time and by taking the imaginary part of the
 equation (\ref{greensep})  we obtain the spectral weight as a function of the
final position of the hole and the frequency $\omega$:
\begin{equation}
A(\omega,R)=\Omega(R) \,\, A^{Free} (\omega,R)
\label{asep}
\end{equation}
The previous expression is exact in one dimension even for this simplified
$t-J_z$ model where the spinon function is particularly simple
$\Omega(R) = \delta_{R,O}$.\cite{parola}
By Eq. (\ref{bethegreen}) spin-charge decoupling is valid in the
Bethe lattice for the single hole problem.

Equation (\ref{asep}) can be considered as a direct
and measurable consequence of
spin-charge separation.
In fact , by measuring the spectral weight for two different positions
of the hole we should get that the ratio:
\begin{equation}
{ A(\omega,R^{\prime}) \over  A(\omega,R) } = {\Omega(R^{\prime} ) \over
\Omega(R) } \,\,=\,\, {\rm independent \,\,\, of \,\,\, } \omega
\label{ratio}
\end{equation}

The ratio $A(\omega,R^\prime) \over
A(\omega,R)$, according to expression (\ref{ratio}),  should be independent of
$\omega$ if spin-charge decoupling occurs. This is unlikely  in
dimension higher than one as the presence of the skeleton paths
 strongly renormalizes the spectral
weight with  the distance of the hole from the origin.\cite{zhongso}
The absence of spin-charge decoupling is particularly evident already
 for the 2C case
(Fig.\ref{SCD}), where $A(\omega,R)/A(\omega,R=O)$ is $\omega$ dependent
 for relatively large $R$ and quite low energy.

 Based on the
 numerical and the analytical results, we conclude that spin-charge
 decoupling can occur mainly in lattices where closed loop  paths are forbidden
by the geometry. In these type of lattices
 the Nagaoka theorem cannot be applied.
We have thus found an interesting relation between the Nagaoka theorem,
spin charge decoupling and presence or absence of skeleton paths in a
given lattice. Of course we cannot rule out more complicated form of
spin-charge
decoupling such as the one discussed in \cite{fabrizio}, or that at finite
 doping the situation may change.\cite{putikka}

\section{Lanczos scheme}

The Lanczos technique is widely used in strongly correlated electron systems.
Contrary to the quantum Monte Carlo technique, it does not suffer the
``fermion sign problem'' or any other  instabilities at low temperature.
Dynamical correlations can be easily obtained using this technique. However,
it has been restricted so far to small systems, typically $4 \times 4$ (for
Hubbard models), or at
most 26 sites (for $t$-$J$ and $t$-$J_z$
models) and 36 sites (for Heisenberg models).\cite{poilblanc}
On the other hand a systematic way
for a finite-size scaling analysis in doped system
such as the  $t$-$J_z$ model is not known yet and
some infinite-system properties are still unclear or
even misleading.

In this section, we will develop a scheme, which allows us
to analyze the infinite system Lanczos
data in a very efficient way, so that we are able to
calculate the spectral function of the $t$-$J_z$ model
with good  accuracy.

The Lanczos method is devised to diagonalize huge Hamiltonian matrix with
dimension $N_h$.
The method starts
with a trial wave function $\phi_T$.
A new basis is generated by Hamiltonian multiplication,
\begin{equation}\label{defsi}
{\bf s}_i= H^i |\phi_T>~~{\rm for}~i=0,1,\dots,n
\end{equation}
 Then  an orthogonal basis
\{ ${\bf e}_i $ \} can be iteratively calculated,
after orthogonalization of the vectors ${\bf s}_i$.
Formally we have
\begin{eqnarray}
b_{i+1}|e_{i+1>}&=&H|e_i>-a_i|e_i>-b_{i}|e_{i-1}> \nonumber\\
a_i&=&<e_i|H|e_i> \nonumber \\
b_{i+1}&=&<e_{i+1}|H|e_{i}>  \label{basis}
\end{eqnarray}
where  $b_0=0$, $|e_0>=|\phi_T>.$
In the Lanczos basis, the Hamiltonian turns  to a tridiagonal form
 where  $a_i$ are the diagonal elements  and $b_i$ the off-diagonal
ones  for $i=1,\dots n$.

For $n=N_h$,  the spectrum of the
tridiagonal matrix coincides with the one of the original Hamiltonian.
Unfortunately the dimension of the Hilbert space is usually given by
 $N_h \sim 10^7 \div 10^9 $, and
it is prohibitive to perform a full diagonalization
 with the available computers.
Nevertheless the ground state and the corresponding energy
converge for relatively small $n\sim 10^2 << N_h$.
This justifies the success of the method which enables to restrict the
diagonalization to a very small basis $n << N_h$ where, according to the
Ritz theorem, the variational principle applies
for all the eigenvalues of the smaller tridiagonal matrix
and in particular for the ground state energy.

\subsection{Lanczos scheme in an Infinite Lattice}

In an infinite system, the Lanczos scheme (\ref{basis}) can be applied
efficiently, provided  there is a proper way to define a simple
{\em finite} basis to represent the vectors $s_n= H^n |\psi_T>$ (\ref{defsi}),
generated  by the iterative application of $H$ to the trial  state.
 This is in fact the case for the
effective $t$-$J_z$ Hamiltonian (\ref{eff}), where ${\bf s}_i$ is represented
by overturned spins on the N\'eel state  localized around the hole.

For any fixed momentum $p$, we start from a N\'eel state with a hole at origin.
Since the $J_z$ term of the Hamiltonian is diagonal,
the only part of the effective $t$-$J_z$ Hamiltonian, relevant to generate
new states, is the kinetic term.
In each multiplication of
the effective Hamiltonian, the hole is translated
to its $z$ nearest neighbors by the translation operator $T_{\tau_\mu}$
(see Fig.(\ref{effp})).
Then the spin exchange operators $\chi_{O,\tau_\mu}$
move the hole  back to the origin,
leaving an overturned spin background and generating $z$ new states.

The possibility to work with a finite basis even in the infinite system was
 first noted by Trugman\cite{trugman}.
In fact the overturned spins are located within a region around the hole
with radius $n$.
We can thus  update only the defects over the N\'eel state, which
are finite at any finite number of multiplications of $H$.
After $n$ steps, the Hilbert space is finite having at most dimension $z^n$.

This exponential growth of the Hilbert space $\sim z^n$
makes the problem intractable even
for relatively small $n$.
Fortunately many of the
generated states appear several times during the expansion of
the Hilbert space, due to the presence of the Trugman-like paths,\cite{trugman}
and also due to the translation symmetry implicitly
exploited by use of the effective Hamiltonian (\ref{eff}).
After all, the dimension of the Hilbert space
turns out to be considerably smaller
than the previous estimate, and in fact it grows
much slower than $z^n$, e.g. $\sim 1.9^n$  for $z=3$. In this way we have
reached  $n=26$  for the 2C case and $n=14$ for the 2D case with an Hilbert
space dimension at most equal to $\sim 12.2 \times 10^6$.

 The {\sl smallest} finite lattice  which contains the full
information of the first exact $n$ Lanczos steps has linear dimension
$2n+1$. Therefore, our results correspond to a $53\times 2$ lattice in the
$2C$ case and to a $27\times27$ in 2D, which is by far larger than the
 size of a typical  finite size Lanczos calculation\cite{poilblanc}.
However the finiteness of $n$ often leads  to appreciable
systematic errors. In these cases we overcome this difficulty by carrying out a
systematic  extrapolation in  $1/n \to 0$, following a scheme which is
analogous to the finite size scaling analysis for finite lattice calculations.

We conclude this section with some technical comments about the algorithm.
 The  basis generated by the iterative application of $H^{eff}_p$ does not
depend on the momentum $p$  of the hole. Hence we only need to generate it
once,  which takes less than two hundred seconds of CPU time on a Cray-C90.
After that,
we can do the usual Lanczos
iterations for fixed parameters  $p$ and $J_z$, which typically takes  $10^2
\div 10^3$ seconds of CPU time on the same computer.

\subsection{Lanczos Spectra Decoding}

With the $n+1$ eigenvalues $E_i$ and eigenfunction $|\Psi_i>$
of the Lanczos matrix truncated after $n$ step,
the spectral weight can be formally calculated as
\begin{equation}
\begin{array}{rcl}
A(k,\omega)&=&{\rm Im} {{1} \over {\pi}} \langle \Psi_T
\vert {{1} \over {\omega -H-i \delta }} \vert \Psi_T \rangle \\
 &=& \displaystyle \sum_{i=0}^n \vert \langle \Psi_i
\vert \Psi_T \rangle \vert ^2 \delta ( \omega - E_i) \label{sweight}
\end{array}
\end{equation}
In the following we assume that the energies $E_i$ are set in ascending order :
$E_{i+1} > E_i$.

As a result  of the finiteness of the restricted Hilbert space,
we get a sum of $\delta$-functions in the spectral weight
at any fixed $n$. Due to the   Lehmann representation of the spectral
function\cite{fetter}
this feature is also present in any finite size calculation.
In this case the thermodynamic limit is obtained by
smoothing the $\delta$-functions in Eq.~(\ref{sweight}) with
Lorenzians of a given small width $\delta$,\cite{poilblanc,dagotto}
\begin{equation}
\delta(\omega-E_i)\,\, \to  \,\, {\rm Im} {{\pi^{-1}} \over
{\omega-E_i-i\delta}}
\label{deltasmooth}
\end{equation}
and then taking the limit  $\delta \to 0$.
For small  finite $\delta$,
 reasonable results can be obtained by the finite size
Lanczos algorithm, provided  that the
resolution of the energy levels becomes  much smaller than $\delta$, i.e.
$n$  large enough but still much less than $N_h$.\cite{poilblanc}

Numerically we have been able to perform
 $n=26$ and $n=14$ Lanczos iterations for the 2C and the 2D lattice
respectively.
 Even though  the ground state
energy is already well convergent, such a small number of Lanczos steps
 is usually far from enough
for a good estimate of the spectral weight.
In fact  by the conventional method of smoothing the
$\delta-$functions described in (\ref{deltasmooth}),  either one miss
 the details of  the spectral weight for large $\delta$ or in the opposite
case  one  gets too rapid oscillations which are obviously
unphysical.\cite{noi}

A more efficient
method for evaluating the  spectral weight was recently introduced by
us.\cite{noi}
  In the following, for reason of completeness  we will give a brief review
of this new method named   Lanczos Spectra  Decoding. In this simple
method, we introduced an interpretation of  the Lanczos
scheme. With this interpretation, the spectral function can be calculated
accurately, efficiently and easily even with a small number of iterations $n$.
As we have seen in fact the
Lanczos  scheme in the infinite system has a computational cost growing
exponentially  with $n$, whereas in any finite size calculation the algorithm
is only  linear in $n$.  We expect therefore that our method is essential in
the
first  case but maybe  helpful only for a small computer-time factor
 in a finite size calculation.

As well known the spectral weight $A(\omega)$ is a distribution that may
be divided into two parts $A(\omega)=A_{coh}(\omega) + A_{incoh}(\omega)$,  a
coherent one $A_{coh}(\omega)$  which contains only $\delta-$function
contributions and an incoherent one $A_{incoh}(\omega) $ which is
a continuous and usually smooth function of $\omega$. The
Lanczos spectra decoding exploits  the smoothness
 properties of $A_{incoh}(\omega)$ in a simple and efficient way.
In the following we therefore assume that the spectral weight is incoherent.
This is not a limitation since coherent peaks can be easily separated out from
the incoherent part by identifying all the quasiparticle weights
  $Z_i=\vert\langle \Psi_i \vert \Psi_T \rangle \vert ^2 $
that remain finite for  $n \to \infty$.

If the spectral weight is incoherent , since by completeness
$\sum\limits_{i=0}^n Z_i=1$, one expects that $Z_i \propto {1\over n}$. Thus
\begin{equation}
Z(\omega)=(n+1) Z_i
\label{zweight}
\end{equation}
may define a smooth function of $\omega$ at the discrete Lanczos energies
$\omega=E_i$.
The full spectral weight is then closely related to the previous function:
\begin{equation}
A(\omega)=Z(\omega) \rho_L(\omega) \label{zrho}
\end{equation}
where  $\rho_L$ represents
the  Lanczos density of states (LDOS) in the restricted  Hilbert space
generated
by the Lanczos algorithm:
\begin{equation}
\rho_L(\omega)= { 1 \over n+1} \sum_{i = 0}^n
\delta (\omega-E_i) \label{rho}
\end{equation}
where the factor ${1\over {n+1}}$ is determined by  the normalization
condition $ \int\limits_{-\infty}^{\infty} d\epsilon \rho_L(\epsilon)=1$.
For $n=N_h$ in a finite system, $\rho_L$ coincides with the actual density of
states of the many body system, but, in an infinite lattice  ( $N_h=\infty$),
this is not generally true,  as it will be shown for the Bethe
 lattice case.

By definition, the number of states $dN$ between  energies
$\epsilon$ and  $\epsilon + d \epsilon$ is given by
\begin{equation}
d N = (n+1) \rho_L(\epsilon) d \epsilon.
\end{equation}
So the Lanczos density of states  can be calculated as
\begin{equation}
\rho_L(\epsilon) = {1 \over n+1} {d N \over d \epsilon}
\end{equation}
Finally , using finite difference instead of differential,
the coarse grained Lanczos
density of states  can be estimated up to order $O({1 \over n^2})$ by
\begin{equation}
\displaystyle \rho_L ( \bar \epsilon_i ) \,=\,{{1} \over{(n+1)(E_{i+1}-E_i)}},
\label{rhod}
\end{equation}
where the energies
\begin{equation}
\bar \epsilon_i={{E_i+E_{i+1}} \over {2}}
\label{emiddle}
\end{equation}
 lie  at the middle of two consecutive eigenvalues.

The function $Z(\omega)$ which is known at energies $E_i$ can be easily
interpolated at the energies $\bar \epsilon_i$ where also $\rho_L$ is known:
\begin{equation}
Z(\bar \epsilon_i) = (Z_{i+1}+Z_i)/2.
\label{zetad}
\end{equation}
If  $Z(\omega)$ is a  twice-differentiable function, eq.~(\ref{zetad})~
 is accurate up to $O({1 \over n^2})$ as well.
 Thus within the same accuracy $A(\omega)=Z(\omega) \rho_L(\omega)$  easily
follows   at the  discrete energies $\omega=\bar \epsilon_i$:
 \begin{equation}\label{lsd}
A(\bar \epsilon_i)= {1 \over 2} \left ( {Z_{i+1}+Z_i \over E_{i+1}-E_i} \right
).
\end{equation}

The knowledge of $A(\omega)$ close to the previous ``special'' points
$\bar \epsilon_i$
can be easily achieved by standard piecewise interpolations or extrapolations.
At the end, we can verify the sum rule $\int A(\omega)d \omega=1  $,
as a check  for the accuracy of the calculation.

\section{Lanczos Spectra Decoding on the Bethe Lattice}

In this section we will show that $Z(\omega)$ and $\rho_L(\omega)$ are well
defined functions and can be calculated exactly in the Bethe lattice.

On the Bethe lattice,
the problem is exactly solvable because the skeleton paths are absent and
the retraceable paths can be summed analytically.
This exact solution has two meaning
for us: (1) as a test for our scheme and assumptions, (2) as an hint
 to interpret the Lanczos scheme for the physical  2C and 2D lattices.


Using the Lanczos basis defined in Eq.(\ref{basis}), the Hamiltonian
has vanishing diagonal part ($a_n = 0$) and
\begin{equation}
\begin{array}{rcl}
b_1 & = & \sqrt{z}  \\
b_n & = & \sqrt{z-1}~~~~~~~~~~~~~{\rm for}\,\,\,\,\,\, n > 1
\end{array}
\end{equation}
This easily follows  since,   on the Bethe lattice, each
multiplication of the Hamiltonian generates $z-1$ new states,
except for the first iteration, which generates $z$ states.
Due to the simplicity of such Lanczos matrix
it is then possible to  compute analytically $Z$ and $\rho_L$
in eqs.(\ref{zweight},\ref{rho}).

In fact the wavefunction components $\psi_i$, $i=0,1,\dots,n$
of an eigenstate with energy $\omega$ satisfy the iterative relation in the
 Lanczos basis:
\begin{eqnarray}
\sqrt{z} \psi_1 &=& \omega \psi_0 ~~~~~~i=0~~~~~~~\nonumber\\
\sqrt{z} \psi_0 + \sqrt{z-1} \psi_2 &=& \omega \psi_1 ~~~~~~i=1~~~~\nonumber\\
(\psi_{i+1}+\psi_{i-1}) \sqrt{z-1} &=& \omega \psi_i ~~~~~~i\ge2\nonumber
\end{eqnarray}
 From the third equation solutions are possible for $\omega^2 <
\epsilon_{BR}^2$ where $\epsilon_{BR}=-2t \sqrt{z-1}$ is the Brinkman-Rice
approximation for the single hole ground state energy. Then the components
of the eigenstates on the Lanczos basis are given by:
$\psi_i={\rm Re}\,\left( A \,\, \lambda^{\displaystyle i-1} \right) $, where
$\lambda=\displaystyle{ \sqrt{\omega^2 -\epsilon_{BR}^2}  -\omega \over
\epsilon_{BR}}$ and the complex number $A$ is obtained by the first two
equations, yielding $\psi_1={\omega \over \sqrt{z} } \,\, \psi_0$ and
$\psi_2=\displaystyle{\omega^2-z \over \sqrt{ z (z-1)} } \,\, \psi_0$. Finally
$\psi_0$ is determined by the normalization condition $\sum\limits_i
\psi_i^2=1$, and the function $Z(\omega)=\lim\limits_{n \to \infty} \psi_0^2
(n+1)$ reads:
\begin{equation}
Z(\omega)= \displaystyle { z^2t^2 [\epsilon_{BR}^2-\omega^2] \over
2[z^4t^4 -(2z+1)t^2\omega^2]}~~~{\rm for}~~~
\omega^2 < \epsilon_{BR}^2
\end{equation}
The Lanczos density of states (\ref{rho}) is then determined by:
\begin{equation}
\rho_L (\omega)= A(\omega) / Z(\omega)
\end{equation}
 where $A(\omega)$ is the
Bethe lattice density of states  in Eq.(\ref{spectralbr}).
In this case, $\rho_L(\omega)$ has singular inverse square root
behavior near the band tail.

By use of the Lanczos Spectra decoding introduced in the previous section we
have obtained a good approximation to the exact solution  with $n
\sim 10$, much less than in the conventional calculation.\cite{noi}

At finite $J_z$,
the only change is with the diagonal part of the Lanczos matrix
\begin{eqnarray}
a_0&=& {J_z \over 4 } z\\
a_i&=& {J_z \over 4} ( 3 z-1 + 2 (z-2) (i-1)) ~~~~ {\rm for} \,\,\, i\ne 0
\end{eqnarray}
This result can be easily
obtained by counting the number of  broken
bonds in the states generated at different Lanczos iterations.
In this formalism we obtain therefore that the motion of a hole in a Bethe
lattice is exactly equivalent to a one dimensional motion of a particle in a
linear  potential, provided we identify the distance of the particle from
the origin with the label $i$ of the Lanczos basis.

By diagonalizing numerically the Lanczos matrix for large $n$ we can easily
 confirm the prediction of the long wavelength hamiltonian (\ref{string}),
namely that  the spectral function is $k-$independent,
and that $A(\omega)$ contains only $\delta-$function peaks.
In fact due to the linear potential {\em all}  of
the  one-hole states are localized states.
It is important to remark that the corrections to the asymptotic $J_z \to 0$
behavior is quite important for the quasiparticle weight $Z$
even for very small $J_z$. For $Z$ the correct linear behavior $\propto J_z$
is found only for $J_z \sim 10^{-3}$ (see Fig.(\ref{stringz})), whereas
the behavior of the ground state energy and the gap are more reasonable.

\section{Results and Discussion}
\subsection{Spectral weight for the $J_z=0$ case}


For $J_z=0$ the spectral weight $A(\omega)$ is an even function of $\omega$ and
in the following we will concentrate on its negative frequency region.
In this model
 we found \cite{noi} a
sharp peak in the spectral weight
 located at an energy close to the retraceable path prediction,
$\epsilon_{BR}=-2 t \sqrt{ z-1} $, and,  a second peak at energy $\sim -t$. In
the 2D case, the spectral weight looks similar, although the first peak
is rather
small. In 1D, the exact BR solution leads only to one peak but with a divergent
 spectral weight $\sim {1 \over \sqrt {\omega-\epsilon_B}} $ at the bottom
 $\epsilon_B$ of the band.
Already in the 2C case such a divergence disappears within the retraceable path
approximation, as well as in our numerical scheme,
which includes all closed-loop paths.

We also found that
the first peak in the spectral function has a remarkable dispersive
feature although the bottom of the spectrum appears  $k$-independent.
The dispersion of the first  peak  is not present neither in 1D or in infinite
dimension\cite{vollhardt}
and the importance to go beyond the retraceable path approximation
is already clear even in 2D.

For the density of states $D(\omega)=\int {d p^d \over (2 \pi)^d}
A(\omega,p)=A(R=O,\omega)$,  our results
present some small
oscillations around the retraceable path analytic solution (\ref{greenbr}).  In
2D however the retraceable path  expression for $z=4$ seems already quite
accurate,
 at least away from the band tails.\cite{noi} All the above results have been
confirmed recently .\cite{mh}

As discussed in the introduction a key question is the
determination of the band edge  energy $\epsilon_B$, i.e. the threshold energy
where the spectral weight begin to vanish.
As a first step we identify  the Lanczos ground state energy $E_n$ for $n
\to \infty$ as $\epsilon_B$, which may or may not converge to the Nagaoka
energy $\epsilon_N$. For instance, by neglecting closed loop paths,
one obtains the Brinkman-Rice energy $\epsilon_{BR}$ as a
variational estimate of $E_{\infty}$. In the retraceable path approximation
$\epsilon_B$ coincides with $E_{\infty}$, as one generally expects.

In order to have an accurate  estimate of $E_{\infty}$, it is useful to have a
guess about  the asymptotic behavior   of the quantity
$\Delta _n = E_n -  E_{\infty}$ for $n\to \infty$.
The way $\Delta _n$ vanishes for $n \to \infty$ is related to the form of
the Lanczos density of states   at low energy. In the Brinkman-Rice case the
exact solution in (\ref{spectralbr}) gives $\rho_L(\epsilon) \sim
(\omega-\epsilon_B)^{-1/2}$. Thus, using Eqs.~(\ref{rhod}), $\Delta_n ^{-1/2}
\sim { 1\over n \Delta _n} $, yielding $\Delta _n \sim {1\over n^2} .$
  For $J_z=0$ the inclusion  of the skeleton paths seems to support a finite
Lanczos density of states (see Fig.(\ref{LDOS})), yielding, by the same
argument
$\Delta _n \sim { 1\over n}.$ We have plotted in Fig.(\ref{ground0})
the estimated ground state energies as a function of $1/n$ for several momenta
for the 2D and the 2C cases. Many of the estimated Lanczos energies -- exact
upper bound of the true ground state energy-- are clearly below $\epsilon_{BR}$
(even for the 2D case also shown in the picture).  Thus,  a previous suggestion
that the one hole energy in a quantum antiferromagnet  should be  close to
$\epsilon_{BR}$ \cite{hasegawa,emery} {\em is not confirmed}  by our numerical
results. In Fig.(\ref{ground0}) -2C case- it is  remarkable that all of
the extrapolated energies are very close to the Nagaoka energy, independent of
the momentum of the hole, i.e. $E_{\infty} = -3 \pm 0.02$.

 The above  results give a robust evidence that
$\rho_{LDOS}$ is finite up to the Nagaoka energy.
Even in this case the spectral weight
$A(\omega)=Z(\omega) \rho_{LDOS}(\omega)$ can in principle
vanish for $\omega > \epsilon_N$ due to the vanishing of the factor
$Z(\omega)$.
This is  the scenario  suggested in  \cite{mh} using the  expansion
for the Green's function  in terms of $K(\omega)$ (\ref{green}).
In the  interval  $K(\omega)  \alpha < 1$ (for $|\omega|/t > {(z-1) \over
\alpha}  + \alpha$  ) when the expansion (\ref{green}) converges the
Green's function is surely real, thus determining  a lower bound for
$\epsilon_B$.
 After some extrapolation it was found in \cite{mh} that $\alpha < z-1$,
i.e.  $\epsilon_B$ should be
higher  than $\epsilon_N$ for the 2C or the 2D case  using the first 18 or 12
coefficients $C_n (R)$ of the Green's function
expansion,  respectively. Although the above analysis is surely correct, the
basic conclusion is affected by systematic errors due to the limited knowledge
of a few coefficients in the expansion.

 As it is shown in
Fig.\ref{weight0}, this scenario looks unlikely within the Lanczos
Spectra decoding method  because $Z(\omega)$ seems to be smoothly connected  to
the Nagaoka energy.  It is clear however that this is not enough for a definite
conclusion.

In order to solve the latter controversy without relying on the Lanczos spectra
decoding, we have reproduced the M\"uller-Hartmann--Ventura expansion (Table
I) and extended it up to the first 26 coefficients for the 2C case.
Since we have generated all of the ${\bf s}_i$, defined in (\ref{defsi}),
 up to $i=n=26$ for the 2C case,
the total number of paths of length $2n$ can be easily calculated as
\begin{equation}
M_n= < {\bf s}_n | {\bf s}_n >
\end{equation}
By proper undressing the retraceable paths, we can convert the total number of
paths $M_n$ to the number of the skeleton paths $C_n$.

As it is shown in table I the M\"uller-Hartmann--Ventura
extrapolation:
\begin{equation}
\label{fitmu}
C_n (R) =C(R) { \alpha^{2n} \over (2n)^{\beta(R)} }
\end{equation}
 used to determine the radius of convergence
$\alpha$ is not stable when large skeleton paths are included.
 For large $n$,
we find that $\alpha$, $\beta(R=O)$ and $C(R=O)$  are
always going up. At $n=26$, $C(R=O)$ becomes $10$ times larger than
their value obtained for $n=18$, while $\alpha$ changes from $1.88$ to $1.91$.
(Table I)

Instead of using the fit (\ref{fitmu}),
we apply the well established  ratio
method, which is well known in the study of critical phenomena.
Using this method we
evaluate the radius of convergence $\alpha$,
and the power law exponent
$\theta$ describing the vanishing of the Green's  function at the band tails,
corresponding to the critical temperature and
to the  conventional critical exponent in the language of critical
phenomena, respectively.\cite{domb}
We define
\begin{equation}
\mu(n)= \sqrt{C_n \over C_{n-1}}
\end{equation}
and the linear intercept at ${1\over n} \to 0$  with two next consecutive
points
\begin{equation}
\mu(n,n-2)={1 \over 2} [n \mu(n)- (n-2) \mu(n-2)].
\label{mudef}
\end{equation}
In fact  $\mu$ is expected to behave as
\begin{equation}
\mu(n)  = \alpha [ 1 + {g \over n} + O({1 \over n^2})]
\end{equation}
where $\alpha$ gives the position of the singularity and $\theta=-(1+g)$ gives
the   exponent of the Green's function at the band tail $G(\omega) \propto
(\omega-\epsilon_B)^{\theta}$.
This kind of analysis is much more stable than the extrapolation
used  in \cite{mh}.
In Fig.(\ref{mh}) the data for the linear intercept $\mu(n,n-2)$ suggests that
 the
$1\over n$ corrections are well behaved, and approximately linear for $n>20$.
By accident this number is very close to the maximum $n$ available in the
previous analysis \cite{mh}, giving further evidence that for this
problem a very large number of coefficients $C_n$  are necessary to obtain
reasonably converged results.

Then, by a linear fit in the region $n >20$, we
obtain  $\alpha = z-1$ within one per cent, both for $R=0$ and for $p=0$,
as an independent check that $\alpha$ should not depend on $R$
 ($C_n(p=0)=\sum_R C_n(R)$).
This is  a clear evidence that
the band edge energy coincides with the Nagaoka energy.

For the 2D case the data  shown in Fig. (\ref{mh2d}) indicate that the
above analysis is poorly converged due to a too much small value of  $n=14$.
This also explains why the extrapolated $n \to \infty $ results for $E_\infty$,
shown in Fig(\ref{ground0}), does not converge to $\epsilon_N$ in 2D.

For the critical exponent $\theta$ we have a much less accurate result
probably because the vanishing of the spectral weight close to $\epsilon_N$
is characterized by overall essential singularities of the type
 $e^{-{1\over \omega -\epsilon_N} }$, as we have suggested in our previous
 paper.\cite{noi}

Here we can give a very simple argument supporting the expected singular
 behavior for $Z(\omega)$.
For the spectral weight the behavior close to $\epsilon_N$
should be the same, because , as we have previously shown,
the Lanczos density of states is approximately constant in this region.

After $n$ Lanczos steps the hole forms a polaron state $|P>$ with maximum spin
but  total $S_z=0$\cite{polaron}
 in a region of linear size $\xi \sim n$, as it is also confirmed by direct
calculation of the spin arrangement around the hole (see next section).
The overlap of the N\'eel state and this trial state
is then approximately given by \cite{polaron}:
\begin{equation}
\label{zanfit}
Z_0 \sim {1 \over { V \choose
V/2} } \to \sqrt{ \pi V \over 2 }  2 ^{-V}
\end{equation}
where $V$ is the
volume of  the polaron equal to $ 2 \xi$ for the 2C case and $\propto \xi^d$
for
general spatial  dimension $d > 1$. The quantity $Z_0 \times n$  according to
the Lanczos spectra decoding method  characterizes the behavior of the smooth
function $Z(\omega)$ at an energy $\omega= \epsilon_N +{\rm const.} / n$.
 Solving for $n$ from the  latter equation and assuming, as we have already
mentioned, that  $\xi \propto n$ we can substitute $V\propto
(\omega-\epsilon_B)^{-d}$ in (\ref{zanfit}) and obtain:
\begin{eqnarray}
A(\omega) \propto Z(\omega) &\propto& (\omega-\epsilon_N)^{-3/2} e^{-{\Delta
\over (\omega -\epsilon_N) }} {\rm ~~~~~~for~ 2C} \label{z2c} \\
A(\omega) \propto Z(\omega) &\propto& (\omega -\epsilon_N)^{-d/2-1} e^{-{\Delta
\over (\omega -\epsilon_N)^d }} {\rm ~~~for~} d > 1 \label{zd}
\end{eqnarray}
where $\Delta$ is an overall constant depending on the dimensionality.
Using the above formulas we have obtained good agreement with numerical results
over a range for $Z$ covering up to two decades (see Fig.(\ref{edg})).

\subsection{Spectral weight for finite $J_z$}
For finite $J_z$ a coherent part shows up in the spectral weight. However,
contrary to the string picture,  only the first few energy levels
contribute  to the spectral weight with true $\delta-$ functions
(See Fig.(\ref{spectral})).
In order to identify these $\delta-$ function contributions we can
check whether the quasiparticle weight $Z_i$ converge to some {\em finite}
value
for  $n \to \infty$, whereas if the energy level $E_i$ belongs to the
incoherent
part $Z(\omega)= (n+1) Z_i$ remains finite for $n\to \infty$.
Another method to distinguish the coherent part from the incoherent one is to
analyze directly the wavefunction components of the eigenstate $\Psi_i$ on the
Lanczos  basis $\{e_j\}$. As we have seen in the previous
section  the label $j$ measures  the length of the overturned
spins  in the state $i$.
The quasiparticle weights $Z_i$ are finite only if
the N\'eel state $e_{j=0}$  will have a non
vanishing  component with the state $\Psi_i$. This obviously occurs
if the eigenstate $\Psi_i$ is ``localized'' in the Lanczos basis even for $n\to
\infty$, otherwise only a probability $\sim {1\over n}$ to be in the N\'eel
state is expected.

Using the above criteria, shown in Fig.(\ref{wf}), we have a
clear evidence of
a single quasiparticle weight for $J_z$ not too large, and an incoherent
part which is rather similar to the $J_z=0$ one. For larger $J_z$ the
incoherent part moves quite fast to higher energies, leaving probably more than
one $\delta-$function contributions to the spectral weight.
We then  conclude that the inclusion of closed loop paths
does not suppress the first quasiparticle weight, but completely washes
out all the quasiparticle excitations at higher energies. Even the few peaks
that appears in the incoherent part cannot be associated to ``string state''
resonances -as suggested by \cite{dagotto,manusakis}- but are rather
similar to the $J_z=0$ ones, where localized string states cannot exist. We
thus
confirm the conclusions of Poilblanc.\cite{poilblanc}

As far as the energy spectrum $E(p)$ is concerned we
obviously found that  the lowest
energy state has  a finite  quasiparticle weight and has momentum
  $p=(0,0)$, instead of
 $({\pi \over 2},{\pi \over 2})$ as
commonly accepted for the $t$-$J$ model. \cite{gros,klr}
This is because in the $t$-$J_z$ model
the spin fluctuations are neglected, while they play an important role
for the energy dispersion  $E(p)$.

Finally for the quasiparticle weight as a function of $J_z$ we can apply the
same argument at the end of the previous section by assuming that $Z$
is basically the overlap of  an $S_z=0$ polaron state of size $\xi$, where
$\xi$ may be roughly identified as the correlation length within the string
picture $\xi \propto J_z^{-1/3}$. We should get essential singularities
like
\begin{eqnarray}
\label{zanfitj}
Z &\propto& J_z^{-1/6} e^{-\Delta J_z^{-1/3}} ~~{\rm for}~~{\rm  2C}
\nonumber\\
  Z &\propto& J_z^{-d/6} e^{{-\Delta \over J_z^{d/3}}} ~~ {\rm for}~~
 d > 1.
\end{eqnarray}
 As in the string picture,  where we could not detect the correct
$Z\propto J_z$ behavior for reasonably small  values of $J_z$ (see
Fig.(\ref{stringz}) ),  we expect that
these kind of singularities are important only at a value of $J_z \sim
10^{-3}$.  For larger values of $J_z$ one obtains a crossover to a
 $J_z^{2/3}$  behavior surprisingly valid for a quite large range of
$J_z$ both in the string picture and in the realistic cases shown in   Fig.(
\ref{stringz},\ref{zfit}).
In the latter picture it is also evident that at some small value of $J_z$
the $Z$ factor should vanish much faster than $J_z^{2/3}$, otherwise we should
get an unplausible critical value of $J_z$, where the quasiparticle weight
vanishes. This at least supports the
behaviour for $Z$ shown in Eq.(\ref{zanfitj}).

\subsection{Ground state properties for small $J_z$}

The accurate determination of the ground state energy in the small $J_z$ region
is important to detect a possible transition between  a uniform
antiferromagnetic state and  a state where the hole fully polarize the spins
in a finite  region of space with size $\xi^d$,  region being
  somehow phase-separated
from the  remaining antiferromagnetic region.
The latter variational
state has an energy  approaching   the  Nagaoka energy for $\xi \to \infty$.
At finite $J_z$, by optimizing the size $\xi$ of the  ferromagnetic
 region,  corrections to the asymptotic Nagaoka energy
scale like  $J_z^{\displaystyle {2 \over d +2}}$.\cite{emery}

Support for a possible transition comes from  the infinite dimension
limit\cite{vollhardt}.  In fact the exact evaluation of the Green's
function, possible in infinite dimension, implies  that
 the lowest energy state contributing to the spectral weight
 is analytically connected at small $J_z$ not to the Nagaoka energy
 $\propto -d$  but to the
much higher Brinkman-Rice energy $\propto -\sqrt{d}$.
If the latter state is identified with the lowest energy state in a uniform
antiferromagnetic phase,
there should be a critical value $J_c$ where the `phase separated' state
 , discussed at the beginning of this section, becomes lower in energy
for $J_z < J_c$.\cite{emery,dagotto}
This is essentially the scenario proposed by Emery et al. for the doped
$t-J$ model.
We will show in the following  a clear numerical evidence that the
above scenario {\em is not confirmed } in the $t-J_z$ model for a single hole,
 at least in the 2C case.

Contrary to the string picture or the infinite dimension limit we see in
Fig.(\ref{grdstex}) that the asymptotic value for the energy is clearly given
by the
Nagaoka energy, although the leading corrections to the energy seem quite well
fitted by the string picture exponent $J_z^{2/3}$.
In fact the diagonal elements of the Hamiltonian in the Lanczos basis
describe approximately a linear potential (Fig.(\ref{po})) as in the
string picture. This is obviously important for the small $J_z$ correction
to the energy.
Moreover in Fig.(\ref{grdstex}) it is shown that
the `phase separated' state is well above the estimated energies
even for very small $J_z$, leaving a possible transition at an
unphysically small value of $J_z \sim 10^{-4}$ for the 2C case.
For the 2D case however we do not have  enough
accuracy as shown in Fig.(\ref{mh2d}) and we cannot exclude a transition at
$J_z \sim 10^{-2}t$.

\subsection{Spin arrangement in the ground state}
Although we have found evidence  that the  Nagaoka energy is the ground state
energy of the  $t$-$J_z$ model for $J_z \to 0$  it is not clear what spin
background is favoured  in this limit.
For instance  we could have that the Nagaoka state is degenerate in the
thermodynamic limit with  an antiferromagnetic state.

In order to solve this issue, we have calculated the hole-spin-spin correlation
function, by measuring  the one involving the spins in the
$z$-direction
\begin{equation}
\begin{array}{rcl}
C_z ^\mu (R_i) &=& N < \psi_p| h_o^\dagger S_{R_i}^z S_{R_i+\tau_{\mu}}^z
|\psi_p>\\&=& <S_O|  S_{R_i}^z S_{R_i+\tau_{\mu}}^z |S_O>
\end{array}
\end{equation}
and its spin rotation invariant version
\begin{equation}
\begin{array}{rcl}
C ^\mu(R_i)& =& N < \psi_p| h_o^\dagger {\bf S}_{R_i} {\bf
S}_{R_i+\tau_{\mu}}|\psi_p>\\&=& <S_O| {\bf S}_{R_i} {\bf
S}_{R_i+\tau_{\mu}}|S_O>
\end{array}
\end{equation}
where $h_o^\dagger$ is the hole creation operator at origin.
These correlation functions measure how the spin background is perturbed
 by the hole.

The symmetrized correlation function $C ^\mu(R_i)$ has been introduced since,
for $J_z \to 0$, the total spin is a well  defined quantum number
and the isotropic hole  spin-spin correlation does not depend on the
polarization of the total spin.
Thus even in the $S_z=0$ sector a polaron solution with maximum spin
leads to a maximum $C(R_i) = {1 \over 4}$, whereas $C_z^{\mu}(R)=0$
for this polarized state since the contribution of the parallel spins
are exactly cancelled by the one of the antiparallel spins
(all these contributions
have the same weight in the polaron solution \cite{polaron}).
 In this
way, we can unambiguously distinguish  the ferromagnetic region with
$C^\mu(R_i)>0$ from   the antiferromagnetic one with $C^\mu(R_i) < 0$.   As
shown in Fig.~\ref{core}, we have found that for large $J_z$ all the spins are
antiferromagnetically correlated in the $z$ direction since the two previous
correlation functions are almost identical. There is a clear evidence of an
antiferromagnetic  correlation which approaches the asymptotic value $C(R) \to
-1/4$ with a correlation length consistent with the one
 shown in Fig.~(\ref{wf}).

For small $J_z$, as expected,  the finite $n$ corrections are important and
tends erroneously to enhance the antiferromagnetism
(see Fig. (\ref{corefit})). Instead by studying the behavior of
 $C(R)$ and $C_z(R)$ as a function of  $1/n$ it is quite clear that
the spins are strongly correlated in the $x-y$ plane and  for $n\to
\infty$ and  $J_z=0$, the hole-spin-spin correlations  seem to approach  the
fully polarized values $C(R)=1/4$ and $C_z(R)=0$, presumably  at any
finite distance from the hole.

\section{Conclusion}

In conclusion
 the physical picture of the one hole ground state in the $t-J_z$ model seems
clear. As we decrease $J_z$,  we approach the Nagaoka state with maximum
spin and  with total $S_z=0$ (which is by the way a conserved quantity)
 due to the proliferation of closed loop paths that strongly  enhance
the ferromagnetic  correlations around the hole in the $x-y$ plane. Thus there
exists a ferromagnetic polaron with a length which diverges for $J_z \to 0$,
but
that is continuously connected to the  antiferromagnetic length obtained at
finite large $J_z$.
In this way the polaron state is somehow similar to the phase separated
variational state\cite{emery}, but we have to emphasize that in our approach
the localized polaron conserves the translational symmetry because it is
defined
after the Galileo transformation (\ref{state}) and consequently it is not
``phase separated''. Moreover the total spin projection on
the $z$-axis vanishes,
as it is conserved locally by the effective Hamiltonian (\ref{eff}).
The energy of this state is much smaller than the ``phase separated''
variational state and represents a more accurate  picture of the single hole
ground state  in the $t-J_z$ model.

We have presented here a successful attempt to go beyond the
retraceable path approximation  and
the string picture for the hole
dynamics in an antiferromagnetic  spin background. A new Lanczos-type
of analysis of the Hamiltonian enabled us to get very accurate results for the
2C problem and qualitatively similar ones for the 2D case.
At $J_z=0$, a clear dispersion of  the main incoherent peak of
$A(k,\omega)$ both for the 2C and the 2D case was found.
Contrary to the prediction of the large spatial dimension we found,
at least for the 2C model, that the bottom of the incoherent band is
dispersionless and coincides with the Nagaoka energy $\epsilon_N$, {\it i.e.}
the minimum possible energy by the Nagaoka theorem.
This resolve a controversy recently proposed by M\"uller-Hartmann and Ventura.
In fact,  based on the Lanczos Spectra Decoding method, we have given an
argument implying that $A(\omega) \propto (\omega-\epsilon_N)^{-d/2 -1} e^{ -
\Delta \over (\omega- \epsilon_N)^d} $ and if for instance $\Delta \propto
(\epsilon_{BR} - \epsilon_N)^d$, we can easily understand why the infinite
dimension limit  gets  no weight for $\omega < \epsilon_{BR}$.
As the dimensionality is increased the tail of the spectral weight below
the Brinkman-Rice energy vanishes exponentially, though remaining always
finite up to the Nagaoka energy in any  finite $d$.

 For finite $J_z$, contrary to the string picture,
we found only one coherent quasi-particle weight and  an incoherent broad
spectrum  at higher energy.
A second  quasiparticle peak may appear in the spectral
function but has always a very small weight.
The possibility of a phase transition as a function of $J_z$ is
not compatible with our numerical results for the ground state energy,
unless for very  small  coupling constant,   and consequently,  the
Emery's argument about the phase separation for small $J$  is found unlikely in
the  $t$-$J_z$ model.

The spin charge decoupling for a single hole
is surely not evident for  short  distance propagations.
However, it still remains open whether the spin charge decoupling
happens asymptotically at large distance, although for the 2C case we have
ruled out this possibility up to a distance of about 10 lattice spacings.

We gratefully acknowledge useful discussions with A. Parola, E. Tosatti and  D.
Poilblanc. This work has been
supported by CNR under Progetto Finalizzato ``Sistemi informatici
e calcolo parallelo''.

\appendix
\section*{  Quasiparticle weight, Green's function and Current operators}

An important quantity to characterize the dynamics of the single
hole    is the so called
quasi-particle weight appearing
as the residue of  a simple pole in the one-hole dynamical Green's function.
This residue  can be alternatively calculated
\cite{anderson2} by means of the overlap of the ground state
$|\psi_p>$ of one hole with
momentum $p$ and the state $c_{p ,\sigma} |H>$, where $|H>$ is
the translation invariant ground state without holes:
\begin{equation}
Z_p \,=\, | <H|c^{\dag}_{p,\sigma_O} |\psi_p> |^2 \,=\,
| <H| S_O> |^2
\end{equation}
 where we have explicitly used that $n_{\sigma,O} |S_O> =
|S_O>$.
For $t>0$, {\it i.e.} positive time, the Green's function is defined as
\begin{equation}
G(p,t)\,=\, -2 i <H| c^{\dag}_{p,\sigma} e^{-i(H -i \delta -E_0) t}
 c_{p,\sigma} |H>
\end{equation}
where $E_0$ is the corresponding energy of the state $|H>$.
Here the factor two comes from the requirement
that $G(p,t\to 0^+) = -2 <H| n_{p,\sigma} |H> = -i$.
The normalized state $|\psi_p> = \sqrt 2 c_{p,\sigma}|H>$ is of the form
(\ref{state}), if we choose
\begin{equation}
S_O^{H}\,=\, \sqrt 2 n_{\sigma,O} |H>.
\end{equation}
Due to the correspondence of eigenstates between $H_p^{eff}$ and $H$,
we can expand $|S_O>$ in terms of eigenstates of $H_p^{eff}$ and easily
check that the propagation of $|S_O>$ with the effective Hamiltonian,
 $|S_O>_t = e^{i H_p^{eff}t} |S_O>$,
 corresponds exactly to the propagation of $\psi_p$ with
the exact $t-J_z$ Hamiltonian and
\begin{equation}
G(p,t)\,=\, -i < S_O^H|e^{-i (H_p^{eff} -i\delta) t} |S_O^H >
\end{equation}
Using that $|S_O^H>= \sqrt 2 n_{i,\sigma}|H>$, that the commutator
$\left[ H_p^{eff},n_{\sigma,O}\right]$ vanishes and
that $G$ does not depend on $\sigma$,
we get, after Fourier transform $G(p,\omega)=\int \limits_{0}^{\infty} d t
\,\,G(p,t)\,\, e^{i \omega t} $,
\begin{equation}
G(p,\omega) \,=\,  <H| {1 \over  \omega + i \delta  -H_p^{eff} }|H>.
\end{equation}

Another important quantity is the current operator, which is useful when we
calculate the transport properties.
On a  discrete lattice, it is defined as\cite{zhang}:
\begin{equation}
j_\mu = \left[ i e_0 t \sum_{R\sigma} c^\dagger_{R\sigma}c_{R+\tau_{\mu}
\sigma} +h.c. \right]
\end{equation}
where $e_0$ is the electron charge.
The matrix elements of the current operator between two one-hole states with
given momentum $p$ define an effective current operator $j^{eff}$ acting on
spin states only:
\begin{equation}  < \psi_p^{S'} |
j_\mu | \psi_p^{S} > = <S'|j_\mu^{eff}|S>
\end{equation}
Analogously to  the calculation shown in Section II, the effective current
operator can be written as
\begin{equation}
j_\mu^{eff}= [-i e_0 t e^{i p \tau_{\mu}}\chi_{O \tau_{\mu}}T_{\tau_{\mu}}
 +h.c.]
\end{equation}

\begin{figure}
\caption{
(a):The ratio of the spectral weight at different  sites for the two chain
lattice.
The spectral weight  as  a function of $R$ was obtained by Fourier transforming
$A(p,\omega)$ obtained by the standard Lanczos Spectra Decoding. Moreover since
at fixed $n$,
$A(R,\omega)$ is exactly zero for $R$ large enough,   only a
finite number of momenta are necessary to implement {\em exactly} the mentioned
Fourier transform (b):The calculated spectral weight for $R=O$.
The solid line is got by summing the spectral weight  for all momenta
verifying  the independent relation $A(R=O,\omega) =\int {d p \over 2 \pi}
A(p,\omega)$.  The data points are calculated directly by the Lanczos Spectra
Decoding using a trial state with a hole localized at the origin, {\it i.e.}
without
using the translation invariance.}
\label{SCD}
\end{figure}
  \begin{figure} \caption{
The application of the effective $t$ Hamiltonian(\protect\ref{heff})
on the trial state. (a):The N\'eel state with one hole.
The hole is located at origin.
(b):The state after the action of $T_{\tau_\mu}$.
$T_{\tau_\mu}$ translates the N\'eel state one lattice space
along the direction  $\mu$. The hole moves to the nearest neighbor.
(c):The final state.
The spin exchange operator moves the hole back to the origin and leaves
an over-turned spin defect in the N\'eel background.}
\label{effp}
\end{figure}
\begin{figure}
\caption{
The single hole quasiparticle weight of the $t$-$J_z$ model on the
Bethe lattice with $z=3$ and $z=4$.
The inset is an expansion of the small $J_z$ region and  the axes have been
scaled by a factor $1000$.
The Lanczos matrix was truncated after  $n=40000$ Lanczos steps, by far enough
to obtain
convergent $n=\infty$ results even for very small $J_z$. }
\label{stringz}
\end{figure}
\begin{figure}
\caption{
The Lanczos density of states  on the Bethe lattice (for $z=3$ and $z=4$),
the 2C and the
2D lattice.
Triangles, squares and circles correspond to the small,  medium
 and large $n$  calculation respectively.
 The continuous lines are the exact results for
the Bethe lattice and guides to the eye for the 2C and 2D lattices}
\label{LDOS} \end{figure}
\begin{figure}
\caption{Plot of the lowest eigenvalues of the $2C$ and 2D model
as a function of $1/n$, the inverse of the Lanczos-iteration number.
For the 2C  case , the wavevector $k$ ranges from
 from $(0,0)$ ( bottom) to $(\pi,0)$  (top) with nine equally spaced values.
For the 2D case  the $k-$path in the magnetic Brillouin zone is
$\Gamma \to M \to X \to \Gamma$, where $\Gamma=(0,0)$ (bottom), $M=(\pi,0)$
(top) and $X=(\pi/2,\pi/2)$.
The horizontal dashed lines denote  the Brinkman-Rice  ground state energies.}
\label{ground0}
\end{figure}
\begin{figure}
\caption{
The expected smooth quantity $Z(\omega)=(n+1) \times Z$ at $J_z=0$ plotted as a
function of the energy for different Lanczos iterations.
The continuous line connects  the $n=26$ data.
By comparing the data at different Lanczos iterations,
$Z(\omega)$ seems to be  non vanishing just above the Nagaoka energy
$\epsilon_N=-3t$ for 2C. The insert is an  expansion of the band edge. }
\label{weight0}
\end{figure}
\begin{figure}
\caption{
The linear intercept $\mu(n,n-2)$ defined in the text (\protect\ref{mudef})
as a function
of ${1 \over n}$ for $p=(0,0)$ and $R=(0,0)$. The dashed lines are linear
extrapolations of the last six points. The two horizontal arrows indicate
the Brinkman-Rice and M\"uller-Hartman-Ventura
estimates. The vertical arrow indicates
the largest $n$ analyzed in \protect\cite{mh}.}
\label{mh}
\end{figure}
\begin{figure}
\caption{Same as in Fig. (\protect\ref{mh}) for the 2D case, the data are
taken from \protect\cite{mh}.
}
\label{mh2d}
\end{figure}
\begin{figure}
\caption{
The behavior of $Z(\omega)$ at band tail for $p=(0,0)$ and $p=(\pi,0)$.
The solid lines are the least square fit by expression  (\protect\ref{zd}).}
\label{edg}
\end{figure}
\begin{figure}
\caption{
The spectral function at $J_z=0.3 t$ and $k=(\pi,0)$ and
$J_z=2t$ and $k=(0,0)$. $Z_1$ and $Z_2$ are the quasiparticle weights for the
lowest two eigenstates. The small value of $Z_2$ compared to $Z_1$ indicates
that  only the lowest weight remains finite for $n\to \infty$, i.e.
contributing to the spectral weight with a true $\delta-$ function.}
\label{spectral} \end{figure}
\begin{figure}
\caption{
The 2C-wavefunction of the single hole in the Lanczos basis for $J_z=2t$,
$p=(0,0)$ and $J_z=0.3t$, $p=(\pi,0)$.
The solid line denotes the ground state, the dotted one the first excited
state and the dashed one the second excited state.
}
\label{wf}
\end{figure}
\begin{figure}
\caption{
Calculated quasiparticle weight $Z$  as a function of $(J_z/t)^{2/3}$.}
\label{zfit}
\end{figure}
\begin{figure}
\caption{
The ground state energy as a function of $J_z^{2/3}$. The data points
refers to  $n=26$ (2C) and $n=14$ (2D).
The continuous lines are  a fit $E=a + b J_z^{2/3} + c J_z$ of the
data.
The dotted line is the energy of the ``phase
separated polaron'' described in the text.
}
\label{grdstex}
\end{figure}
\begin{figure}
\caption{
The  diagonal matrix elements of the Hamiltonian in the Lanczos basis.
The solid line refers to  the two-chain lattice  and the dashed line to the
corresponding  Bethe lattice  with $z=3$.}
\label{po}
\end{figure}
\begin{figure}
\caption{
The hole spin-spin correlation function $C^{\mu}(R)$ and $C^{\mu}_z(R)$ with
$\tau_{\mu}=(1,0)$. For large $J_z$ $C(R)$ is
purely antiferromagnetic. However  for vanishing (or small)  $J_z$, the
ferromagnetic component in the $x-y$ plane is important.}
\label{core}
\end{figure}
\begin{figure}
\caption{
The extrapolation $n \to \infty$ of $C^{\mu} (R)$ and $C_z^{\mu} (R)$ for the
 three lattice  sites  closest  to the hole at $J_z=0$.
The horizontal line is the value of $C^\mu(R)$ for the fully polarized Nagaoka
state. }
\label{corefit}
\end{figure}
\widetext

\begin{table}
\caption{The non vanishing coefficients $C_n(R)$  of the $t$-$J_z$ Hamiltonian
on the 2C lattice.
Only those coefficients  which are not included in the
M\"uller-Hartmanni-Ventura's table
are shown.
The notation is similar to the one  in ref. \protect\cite{mh},
i.e. $d^2=|R|^2$ and
$n_{d^2}$ is the number of skeleton paths at a given distance and for a given
direction (note that there is an extra  factor  two   for $R\ne O$
if we do not distinguish the two
possible directions as in ref.  \protect\cite{mh}). The data for $n=26$ are
 accurate up to $\protect\pm 5$.}

\small{
\begin{tabular}{cccccccc}
\hline
\hline
$2 n$ & $C_n(d^2=0)$ &$C_n(d^2=2)$ &$C_n(d^2=4)$ &$C_n(d^2=10)$ &
$C_n(d^2=16)$ &$C_n(d^2=26)$ & \\
\hline
          32 &
      193448 &
      152314 &
      108676 &
       70960 &
       39081 &
       17614 &
 \\
 \hline
          34 &
      590154 &
      472488 &
      340675 &
      223698 &
      128823 &
       58446 &
 \\
 \hline
          36 &
     1824844 &
     1471492 &
     1068182 &
      708496 &
      414196 &
      196574 &
 \\
 \hline
          38 &
     5677040 &
     4609274 &
     3385018 &
     2264848 &
     1348874 &
      670086 &
 \\
 \hline
          40 &
    17818480 &
    14539266 &
    10760828 &
     7287326 &
     4397638 &
     2245908 &
 \\
 \hline
          42 &
    56220728 &
    46154304 &
    34459409 &
    23573566 &
    14437674 &
     7572526 &
 \\
 \hline
          44 &
   178693158 &
   147425926 &
   110826815 &
    76581474 &
    47561630 &
    25474418 &
 \\
 \hline
          46 &
   570790364 &
   473551402 &
   358473393 &
   249911680 &
   157303528 &
    85932926 &
 \\
 \hline
          48 &
  1834737522 &
  1529492974 &
  1164976270 &
   819090516 &
   521903825 &
   290492088 &
 \\
 \hline
          50 &
  5926011194 &
  4963905566 &
  3804186739 &
  2695446152 &
  1737377480 &
   983973358 &
 \\
 \hline
          52  &
  19240493885 &
  16187397249 &
  12476330859 &
  8905934658  &
  5800668507  &
  3338934862  &
 \\
 \hline
 \hline

\end{tabular}

\bigskip
\bigskip
\bigskip

\vskip 3cm

\begin{tabular}{ccccccccc}
 \hline
 \hline
$2 n$ &$C_n(d^2=36)$ & $C_n(d^2=50)$ &$C_n(d^2=64)$
&$C_n(d^2=82)$ &$C_n(d^2=100)$ &
$C_n(d^2=122)$ &$C_n(d^2=144)$ &
$C(p=0)$ \\
\hline
          32 &
        5714 &
        1140 &
           0 &
           0 &
           0 &
           0 &
           0 &
      984446
 \\
 \hline
          34 &
       19510 &
        4294 &
         534 &
           0 &
           0 &
           0 &
           0 &
     3087090
 \\
 \hline
          36 &
       70917 &
       18258 &
        2981 &
           0 &
           0 &
           0 &
           0 &
     9727036
 \\
 \hline
          38 &
      251056 &
       67790 &
       12460 &
        1190 &
           0 &
           0 &
           0 &
    30898232
 \\
 \hline
          40 &
      883914 &
      258922 &
       55639 &
        7634 &
           0 &
           0 &
           0 &
    98692630
 \\
 \hline
          42 &
     3126095 &
      968326 &
      222326 &
       35058 &
        2661 &
           0 &
           0 &
   317324618
 \\
 \hline
          44 &
    10934279 &
     3569044 &
      886787 &
      164380 &
       19237 &
           0 &
           0 &
  1025581138
 \\
 \hline
          46 &
    38263835 &
    13147300 &
     3468307 &
      697614 &
       96205 &
        5944 &
           0 &
  3332494632
 \\
 \hline
          48 &
   133406362 &
    47855170 &
    13327978 &
     2901958 &
      472897 &
       47830 &
           0 &
  10882673258
 \\
 \hline
          50 &
   464563759 &
   173622392 &
    50912034 &
    11813430 &
     2117499 &
      258896 &
       13277 &
  35702392358
 \\
 \hline
          52 &
  1616935261 &
   626834742 &
   192138728 &
    47077158 &
     9171222 &
     1332690 &
      117733 &
 117646241223
 \\
 \hline
 \hline

\end{tabular}
}
\label{table1}
\end{table}

\narrowtext

\begin{table}
\caption{Fit of the coefficients $C_n(R)$ for $R=O$ using the ansatz
(\protect\ref{fitmu}), in the given intervals of $n$ given in the rightmost
column.}

\begin{tabular}{cccc}
\hline
\hline
$\alpha$ & $\beta_{R=O}$ & $C_{R=O}$ & $2 n$ \\
\hline
 1.88136&     2.39084&     1.26191&   22-36 \\
 1.88356&     2.42844&     1.38458&   22-38 \\
 1.89515&     2.63760&     2.34793&   22-40 \\
 1.89049&     2.54888&     1.86730&   22-42 \\
 1.89824&     2.70449&     2.81409&   22-44 \\
 1.89667&     2.67137&     2.57441&   22-46 \\
 1.90428&     2.83985&     4.08250&   22-48 \\
 1.90575&     2.87370&     4.48588&   22-50 \\
 1.91050&     2.98860&     6.20875&   22-52 \\
\hline
\hline
\end{tabular}

\vskip 3cm
\begin{tabular}{cccc}
\hline
\hline
$\alpha$ & $\beta_{R=O}$ & $C_{R=O}$ & $2 n$ \\
\hline
 1.88136&     2.39084&     1.26191&     22-36  \\
 1.88399&     2.43662&     1.41397&     24-38  \\
 1.89752&     2.68445&     2.65495&     26-40  \\
 1.89228&     2.58627&     2.06374&     28-42  \\
 1.89925&     2.72657&     2.98843&     30-44  \\
 1.89721&     2.68362&     2.66328&     32-46  \\
 1.90523&     2.86235&     4.34957&     34-48  \\
 1.90657&     2.89433&     4.75851&     36-50  \\
 1.91148&     3.01404&     6.68424&     38-52  \\
\hline
\hline
\end{tabular}
\label{table2}
\end{table}

\begin{references}
\bibitem{polaron} The  state $<P|$ with maximum spin and $S_z=0$
has the same overlap with  all  the  possible states of the Hilbert space with
$S_z=0$, thus  in particular $<P|N>^2 = {1 \over { V \choose V/2 } } $, where
$V=\xi^d$  is the volume occupied by the polaron, and the binomial
coefficient $V \choose V/2$ is the number of ways to put $V/2$ spin up and
$V/2$ spin down in a region of $V$ spins.
\end{references}
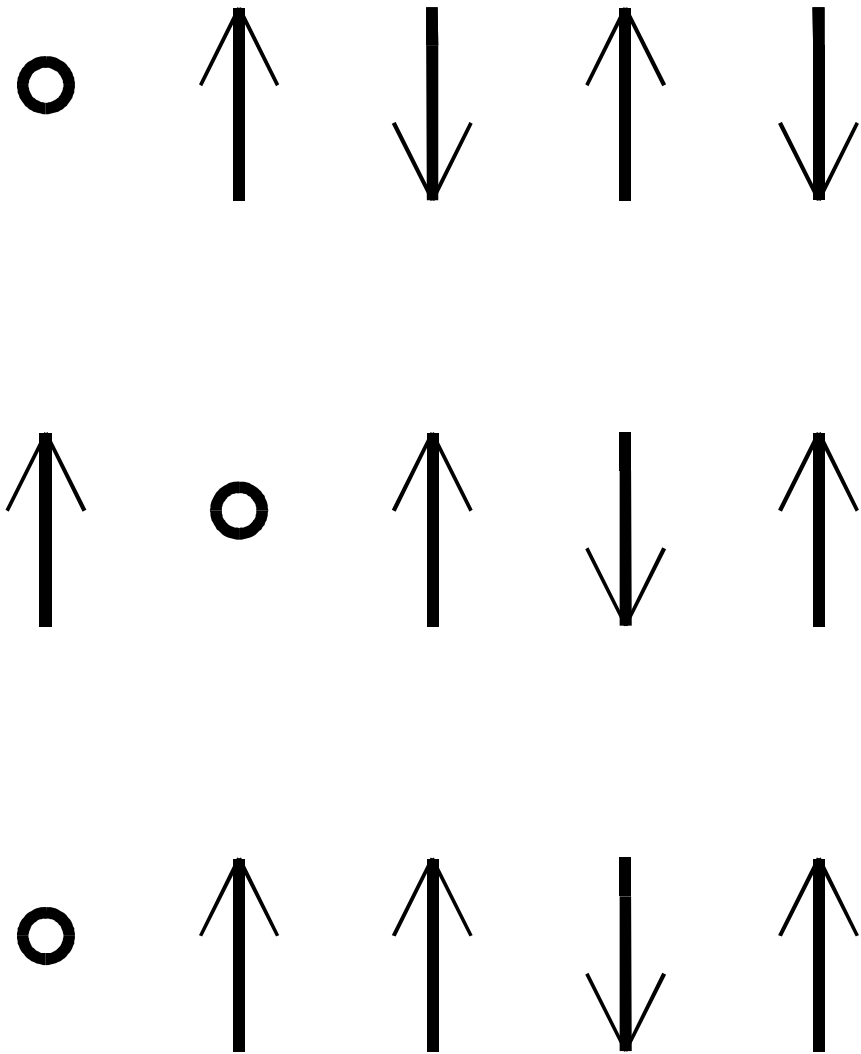
\end{document}